\documentclass[]{aastex631}
\usepackage{appendix}
\usepackage{datatool,booktabs}
\usepackage{subfigure} %

\usepackage{amsmath}

\def\aap{Astron.\ Astrophys.\ }

\def\apjl{Astrophys.\ J.\ Lett.\ }
\def\apjs{Astrophys.\ J.\ Supp.\ }

\newcommand{\zcr}[1]{\textcolor{black}{   #1}}
\newcommand{\zcrr}[1]{\textcolor{black}{  #1}}

\begin{document}

\title {Examining Temporal Characteristics of GCRs in the AMS-02 era} 

\author{Cheng-Rui Zhu}
\email{zhucr@anhu.edu.cn}
\affiliation{ Department of Physics, Anhui Normal University, Wuhu, Anhui, 241000, China}

 \author{Yu-Lu Liu}%
 \affiliation{ Department of Physics, Anhui Normal University, Wuhu, Anhui, 241000, China}

\author{Mei-Juan Wang}%
 \affiliation{ Department of Physics, Anhui Normal University, Wuhu, Anhui, 241000, China}
 
 \author{Tian-Hao Liu}%
 \affiliation{ Department of Physics, Anhui Normal University, Wuhu, Anhui, 241000, China}
  \author{Yi-Hua Zhu}%
 \affiliation{ Department of Physics, Anhui Normal University, Wuhu, Anhui, 241000, China}
 
\author{Kai-Kai Duan}
\affiliation{Key Laboratory of Dark Matter and Space Astronomy, Purple Mountain Observatory, Chinese Academy of Sciences, Nanjing 210023, China }


\date{\today}

\begin{abstract}

The energy spectra of Galactic cosmic rays (GCRs) serve as critical probes of their astrophysical origins and propagation mechanisms through the interstellar medium. Among the key physical processes shaping the observed spectra, solar modulation, driven by turbulent heliospheric magnetic fields and the variability of the solar wind, plays a dominant role in modifying GCR fluxes within the inner heliosphere. The recent release of time-dependent flux measurements for He, Li, Be, B, C, N, and O by AMS-02 provides unprecedented precision to investigate  solar modulation phenomena. In this study, we demonstrate that Li, Be, B, C, N, and O nuclei exhibit solar modulation parameters consistent with those of protons (p) and He within a modified force-field approximation (FFA) framework incorporating rigidity-dependent modulation potentials. By positing that GCRs with the same charge-sign sharing the solar modulation parameters, we further forecast daily fluxes of  Ne, Mg and Si from 2011 to 2020.


\end{abstract}



\section{Introduction}

Galactic cosmic rays (GCRs) represent a captivating realm of high-energy astrophysics. They are charged, highly energetic particles that eject from cosmic accelerators, prominent among which are supernova remnants. After escape from accelerators, they propagate ceaselessly throughout the expanse of the Milky Way galaxy~\citep{1998ApJ...493..694M}. The moment these GCRs penetrate the heliosphere, a dynamic and complex interaction unfolds. They come under the influence of the outward - moving magnetized solar wind plasma, a phenomenon known as solar modulation~\citep{Potgieter2013}.
The study of solar modulation is of paramount importance, as it serves as a key to unlocking the mysteries surrounding GCRs. By delving into solar modulation, we can gain profound insights into the very nature of GCRs, encompassing their distant origins~\citep{2013A&ARv..21...70B} and the intricate propagation mechanisms that govern their movement within the galaxy~\citep{2007ARNPS..57..285S}. Moreover, solar modulation holds a significant place in the search for dark matter. It has a direct impact on the low-energy fluxes of antiprotons and antideuterons, which are crucial observables in the quest to understand the elusive nature of dark matter~\citep{2012CRPhy..13..740L,Yuan:2014pka,2017PhRvL.118s1101C,PhysRevLett.129.091802,2022PhRvL.129w1101Z}.

In recent years, the field of GCR research has witnessed remarkable experimental breakthroughs. Data from experiments, including Voyager, AMS-02, PAMELA, and DAMPE, have revolutionized our understanding of solar modulation and GCR physics~\citep{2013Sci...341..150S,2011Sci...332...69A,2017PhRvL.119y1101A,cite-key}. Of these, Voyager-1 and Voyager-2 stand out as true pioneers. They are the only spacecraft to have successfully crossed the heliospheric boundary, venturing into the uncharted territory beyond the influence of the solar wind~\citep{2013Sci...341..150S,voyager-2}. This extraordinary feat has enabled them to provide direct measurements of the Local Interstellar Spectrum (LIS) in the energy range of a few to hundreds of MeV/nucleon, offering a unique detection of GCRs before they are affected by solar modulation.

The AMS-02 collaboration has also made significant contributions. Their publication of high-precision cosmic ray spectra~\citep{AGUILAR20211} and the detailed documentation of the evolution of some cosmic ray fluxes over time~\citep{PhysRevLett.121.051101,PhysRevLett.121.051102} have opened up new avenues for studying solar modulation in GCRs. These data provide a rich resource for researchers to test theoretical models and refine our understanding of the complex processes at play.

\zcr{To comprehend the propagation of GCRs within the heliosphere, the Parker's transport equation (TPE)~\citep{1965P&SS...13....9P} is used to describe the propagation. This  equation provides a theoretical framework to describe the complex interplay between GCRs and the heliospheric environment. Solving the TPE is no simple feat, and researchers have employed a variety of methods, ranging from numerical simulations that can handle the complexity of the equation with high precision to analytical approaches that offer elegant and intuitive solutions. Among these methods, the force field approximation (FFA) has emerged as a popular choice~\citep{1967ApJ...149L.115G,1968ApJ...154.1011G}. The widespread utilization of the FFA lies in its simplicity, which allows for relatively straightforward calculations while still being sufficient to explain a wide range of observational data.}

However, the FFA, despite its simplicity and wide-spread use, has its limitations. It struggles to fully explain the observed fluxes of cosmic rays and their long-term variations when using homogeneous parameters~\citep{2021ApJ...921..109S,PhysRevD.109.083009}. To address these shortcomings, several innovative methods have been proposed to modify the FFA. These modified approaches aim to provide a more accurate description of cosmic ray fluxes, taking into account the complex and dynamic nature of the heliospheric environment ~\citep{2016ApJ...829....8C,2017PhRvD..95h3007Y,Zhu:2020koq,2021ApJ...921..109S,PhysRevD.106.063021,PhysRevD.109.083009,zhu2024}. \zcr{Here we employ the modified FFA parameterisations proposed by \cite{zhu2024} and \cite{PhysRevD.109.083009}, which introduce a rigidity‑dependent modulation potential via a sigmoid function and a logarithmic form, respectively.}

Recently, AMS-02 released time-dependent flux data for He, Li, Be, B, C, N, and O spanning from 2011 to 2022 \citep{PhysRevLett.134.051001}, 
offering new avenues to study solar modulation. 
\zcr{In this study, we assume that all positively charged GCRs share identical solar modulation parameters, based on the charge‑sign dependence of solar modulation—drift effects act oppositely on particles of opposite signs while same‑sign particles experience consistent behaviour. This assumption is supported by \cite{2022JCAP...10..051C,PhysRevD.106.063021,Zhu_2025}, and the recent AMS‑02 findings \citep{PhysRevLett.134.051001}, which show that positively charged nuclei fluxes exhibit similar time variations and no significant modulation differences related to the mass‑to‑charge ratio (A/Z). Under this assumption, we derive the LIS of Li, Be, B, C, N, O, Ne, Mg, and Si using cubic spline interpolation \citep{2016A&A...591A..94G,2017AdSpR..60..833G,Zhu:2018jbk}, as detailed in Section 2.4.}
\zcr{Subsequently, we calculated the daily fluxes of Li, Be, B, C, N, O, Ne, Mg, and Si from 2 to 60 GV. This calculation uses the derived LIS and the solar modulation parameters obtained by \cite{Zhu_2025}, who fit the AMS‑02 p and He daily fluxes using the modified FFA models of \cite{zhu2024} and \cite{PhysRevD.109.083009}.} Remarkably, most of the predicted daily fluxes lie within the 2 \(\sigma\) confidence intervals across all measured rigidities. With the same methods we have success forecasted the daily fluxes of D, $^3$He, $^4$He \citep{zhu2025probing} and antiprotons \citep{zhu2025antip}. The derived high-resolution flux profiles establish critical benchmark datasets for future GCR detection.

\section{Methodology}\label{sec:meth}

\subsection{Solar Modulation}

The existence of heliospheric magnetic field carried by solar winds causes the modulation of GCRs as they enter the heliosphere, resulting in suppressed fluxes of CRs. This phenomenon is known as solar modulation, and its effects are more pronounced at particle energies below 30 GeV/n~\citep{Potgieter2013}. The basic transport equation was first derived by Parker~\citep{1965P&SS...13....9P}. 3D time-dependent self-consistent modelling is a full solution to the CRs transport problem in the heliosphere, however it is a complicated task and requires a very large amount of computation. The FFA is a simplified method to solve this problem. Usually, the FFA requires the  quasi-steady changes,  spherical symmetry, etc, \zcr{which} are apparently invalid for short time scales. In fact, they are not fully valid even for the regular condition \citep{2003JA010098}. However, the force-field formalism was found to provide a very useful and comfortable mathematical parametrization of the GCR spectrum even during  a major Forbush decrease (FD), irrespective of the (in) validity of physical assumptions behind the force-field model. So one can still benefit from the simple parametrization  offered by the FFA for practical uses other than studying the physics of solar modulation, even if the FFA is not a physically motivated solution to the solar modulation problem at daily time scales \citep{USOSKIN20152940}.
In this model, the top-of-atmosphere (TOA) flux is related to the local interstellar spectrum (LIS) flux as follows \citep{1968ApJ...154.1011G,2011JGRA..116.2104U}:  
\begin{equation}\label{force_filed}
J^{\rm TOA}(E)=J^{\rm LIS}(E+\Phi)\times\frac{E(E+2m_p)}
{(E+\Phi)(E+\Phi+2m_p)}, 
\end{equation}
where $E$ is the kinetic energy per nucleon, $\Phi=\phi\cdot Z/A$ with 
$\phi$ representing the solar modulation potential,
$Z$ and $A$ are the atomic number and mass number of the cosmic ray particle, respectively,
$m_p=0.938$ GeV is the 
proton mass, and $J$ denotes the differential flux of GCRs. The sole parameter in the FFA is the modulation potential $\phi$.


\subsection{Modified Force-Field Approximation}

The force-field model assumes a quasi-steady-state solution to Parker's transport equation. However, observational GCR fluxes exhibit 11-year variations linked to solar activity. To account for this, a time-series of $\phi$ at different epochs is used to describe the data. Since a single parameter cannot adequately fit the monthly cosmic ray fluxes, a rigidity-dependent solar modulation potential is required~\citep{SIRUK20241978}.  The rigidity-dependence could be caused by the diffusion coefficients or the drift coefficient of GCRs within heliosphere.

\zcr{In our prior work \citep{Zhu_2025}, we performed a comparative study of several modified FFA models by fitting them to the daily proton and helium fluxes measured by AMS-02. Among the various models tested, both Zhu's model and Long's model yielded the best and most comparable results, with very small chi2/d.o.f values. This demonstrated that both modified model provide an equally excellent description of the daily p and He data. Therefore, in the present work, we adopt these two models to ensure consistency with our previous findings and to demonstrate that our predictions for heavy nuclei are robust and not strongly dependent on the specific choice of the modulation potential parametrization. }

\zcr{The first modified force-field approximation model is the Zhu's model from \cite{zhu2024}.}
This approach is based on the model developed by \cite{2016ApJ...829....8C}, but replaces its transition function with a sigmoid function to achieve a smoother transition. The solar modulation potential in this model is defined as:
\begin{equation}\label{Zhu}
\phi (R)_{Zhu} = \phi_l +\left (\frac{\phi_h-\phi_l}{1+e^{(-R+R_b)}} \right ),
\end{equation}
where $\phi_l$ and $\phi_h$ are the modulation potentials for low and high energies, respectively, $R$ is the rigidity, $R_b$ is the break rigidity and $e$ is the natural constant. 
The sigmoid function is employed to ensure a smooth transition between $\phi_l$ and $\phi_h$, which is necessary to model the gradual change in modulation potential across different energy ranges. The parameters $\phi_l$, $\phi_h$, and $R_b$ are free and determined through fitting.

\zcr{The second modified force-field model, taken from \cite{PhysRevD.109.083009} and built upon the framework of \cite{Kuhlen_2019}, is given by }
\begin{equation}\label{Long}
 \phi (R)_{Long} = \phi_0 + \phi_1 ln(R/R_0), 
\end{equation}
with 
\begin{equation}\label{force_filed3}
\begin{aligned}
J^{\rm TOA}(E)= & J^{\rm LIS}(E+ \Phi(R)) \\
&\times \frac{E(E+2m_p)}{(E+\Phi(R))(E+\Phi(R)+2m_p)} \\ 
&exp(-g\frac{10R^2}{1+10R^2}\phi(R))).
\end{aligned}
\end{equation}
\zcr{Here, $\phi_0$ is the normalization of $\phi(R)$ at $R_0$=1 GV, $\phi_1$ is the slope of $\phi(R)$ with respect to $\ln R$, and $g$ is a scaling factor related to the magnitude of the magnetic field  and solar wind velocity, which controls the strength of the exponential suppression term at high rigidities introduced to improve the model's performance.} In this model, 
$\phi_0$, $\phi_1$ and $g$ are the free parameters to be fitted with $\Phi(R) = \phi(R) \cdot Z/A $.  

\subsection{Markov Chain Monte Carlo (MCMC)}

We fit the two  solar modulation $\phi(R)$ with three free parameters. The $\chi^2$ statistics is defined as
\begin{eqnarray}
\chi^2=\sum_{i=1}^{m}\frac{{\left[J(E_i;\phi(R))-
J_i(E_i)\right]}^2}{{\sigma_i}^2},
\label{eq:chi2}
\end{eqnarray}
where $J(E_i; \phi(R))$ is the expected modulated flux, $J_i(E_i)$ is the measured flux, and $\sigma_i$ is the measurement error for the $i$-th data bin, with $E_i$ representing the geometric mean of the bin edges. The errors $\sigma_i$ include both statistical and systematic uncertainties, ensuring a robust fit.

To minimize the $\chi^2$ function, we employ the MCMC algorithm within a Bayesian framework. The posterior probability of the model parameters $\theta$ is given by:  
\begin{eqnarray}
p(\boldsymbol{\theta}|{\rm data}) \propto {\mathcal L}({\rm data}|\boldsymbol{\theta})
p(\boldsymbol{\theta}),
\end{eqnarray}
where ${\mathcal L}({\rm data}|\boldsymbol{\theta})$ is the likelihood function and 
$p(\boldsymbol{\theta})$ is the prior probability, chosen to reflect physical constraints on the parameters.

The MCMC driver, adapted from {\tt CosmoMC}~\citep{2002PhRvD..66j3511L,Liu_2012}, uses the Metropolis-Hastings algorithm. Starting from a random point in the parameter space, the algorithm proposes new points based on the covariance of the parameters, which is estimated from preliminary fits. The acceptance probability for a new point is defined as:  

\begin{eqnarray}
P_{\text{accept}} = \min\left(1, \frac{P(\theta_{\text{new}} | \text{data})}{P(\theta_{\text{old}} | \text{data})}\right). 
\end{eqnarray}  
If accepted, the process repeats from the new point; otherwise, it reverts to the previous point. This iterative process ensures thorough exploration of the parameter space, yielding robust estimates of the modulation parameters.
For more details about the MCMC one can refer to~\citep{MCMC}.

\subsection{LIS of  Li, Be, B, C, N, O, Ne, Mg and Si}
\label{sec:lis}

Typically, power-law or broken power-law functions are used to fit GCR data~\citep{2014A&A...566A.142Y}. However, if the observational data span a sufficiently wide energy range, a non-parametric method such as spline interpolation can be employed~\citep{2016A&A...591A..94G,2017AdSpR..60..833G,Zhu:2018jbk}. Spline interpolation constructs a smooth function passing through a series of points using piecewise polynomial functions. Here, we use cubic spline interpolation, with the highest polynomial order set to three, to ensure a smooth and accurate representation of the \zcr{LIS~\citep{2016A&A...591A..94G,zhu2024}.}

\begin{figure*}
    \centering
    \includegraphics[scale=0.6]{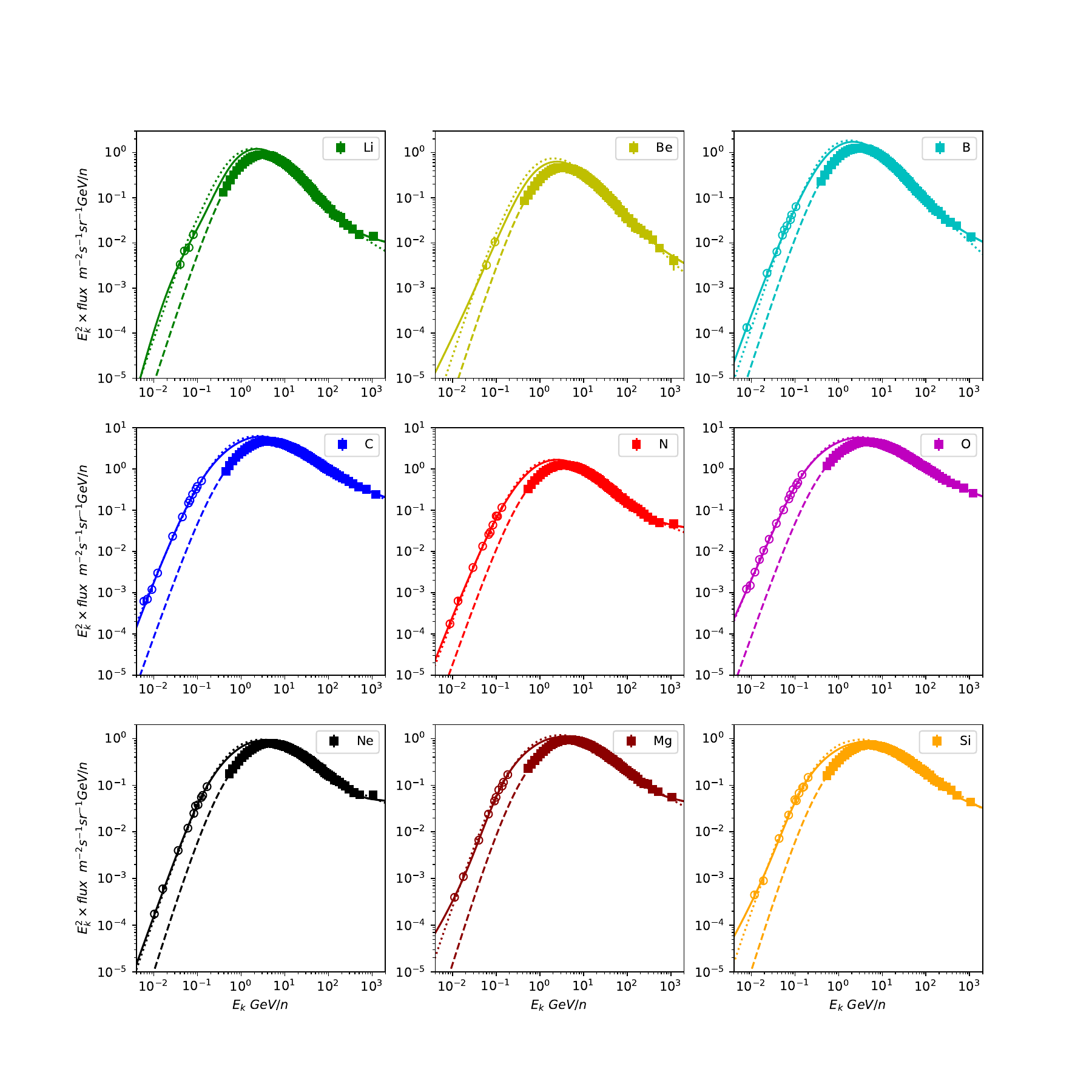}
    \caption{ LIS of Li, Be, B, C, N, O, Ne, Mg and Si. Square points denote AMS-02 data, while circular points represent Voyager observations. Solid lines indicate the LIS, and dashed lines signify the TOA. For comparison, the LIS from the Helmod model \citep{Boschini_2020} is also shown as dotted lines.}
    \label{fig:LIS}
\end{figure*}

\zcr{To derive the LIS for Li, Be, B, C, N, O, Ne, Mg, and Si, we employ the FFA as the modulation model to connect the observed top-of-atmosphere fluxes with the underlying LIS. The modified FFA is not adopted in this step, as it would introduce degeneracies among the fitting parameters, as demonstrated in \cite{zhu2025antip}. In this procedure, we assume zero solar modulation potential for Voyager data, which provide measurements in the local interstellar medium. For the AMS-02 data from the same time period \citep{AGUILAR20211}, we adopt the same modulation potential for all positively charged nuclei as that determined for protons and helium in \cite{zhu2024}, with a value of 0.477 GV. The charge-to-mass ratios are taken as Z/A = 3/6.5 for Li, 5/10.7 for B, 7/14.5 for N, and 1/2 for the remaining species. ACE data are excluded from the fitting procedure, as their rigidity range is too low and the associated modulation potential differs significantly from that at AMS-02 energies due to its rigidity dependence, whereas our focus is on AMS-02 measurements. We employ the MCMC method to determine the best-fit LIS by minimizing the $\chi^2$ statistics defined in Equation \eqref{eq:chi2}. The resulting LIS are shown in Figure \ref{fig:LIS}, alongside the Helmod model predictions \citep{Boschini_2020} for comparison. Overall, our LIS are in good agreement with the Helmod model, with minor discrepancies for Li and Be, which can be attributed to the sparse Voyager observations at low rigidities.}

\section{results and discussion }


\zcr{Our methodological framework rests on two key pillars: (i) the derivation of reliable LIS for heavy nuclei, and (ii) the assumption that all positively charged GCRs share identical solar modulation parameters. To validate this approach, we first benchmark our predictive capability against previously published AMS-02 measurements. In our earlier work \citep{zhu2025antip}, we predicted daily antiproton fluxes using the same framework; the subsequently released AMS-02 data \citep{PhysRevLett.134.051002} confirmed that our predictions lie within the 
1$\sigma$ confidence interval across all measured rigidities. This successful validation provides strong confidence in extending the same methodology to the present study of heavy nuclei.}

\zcr{The solar modulation parameters employed here are derived from daily proton and helium energy spectra \citep{Zhu_2025}, utilizing two distinct modified force-field approximation models: the sigmoid-type model proposed by \cite{zhu2024} and the logarithmic model developed by \cite{PhysRevD.109.083009}. For comparison, we also include results obtained with the standard force-field approximation, using the constant modulation potential derived in the same work \citep{Zhu_2025}. By combining these independently obtained modulation parameters with the LIS reconstructed in Section \ref{sec:lis}, we compute the daily fluxes for Li, Be, B, C, N, O, Ne, Mg, and Si over the period 2011–2020. This prediction constitutes a stringent test of the underlying assumption that all positively charged nuclei experience identical modulation.}

\begin{figure*}
    \centering
    \includegraphics[scale=0.7]{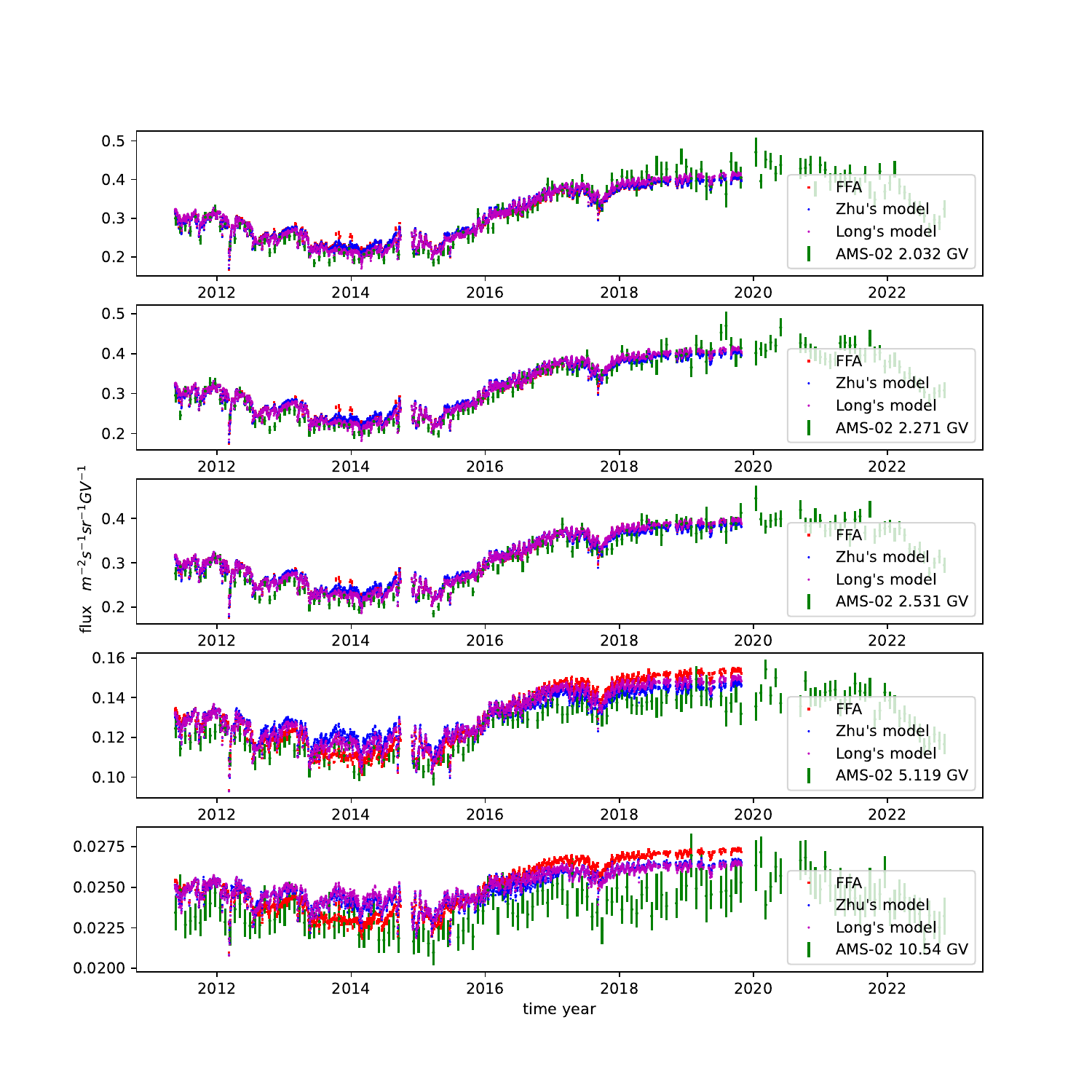}
    \caption{FFA model (red points), Zhu's model (blue points) and Long's model(magenta)  prediction of Li fluxes comparing to the AMS-02 data (green points data) at Rigidities = 2.032 GV, 2.271 GV, 2.531 GV and 5.119 GV and 12.49 GV (from top to bottom). }
    \label{fig:Li}
\end{figure*}

\begin{figure*}
    \centering
    \includegraphics[scale=0.7]{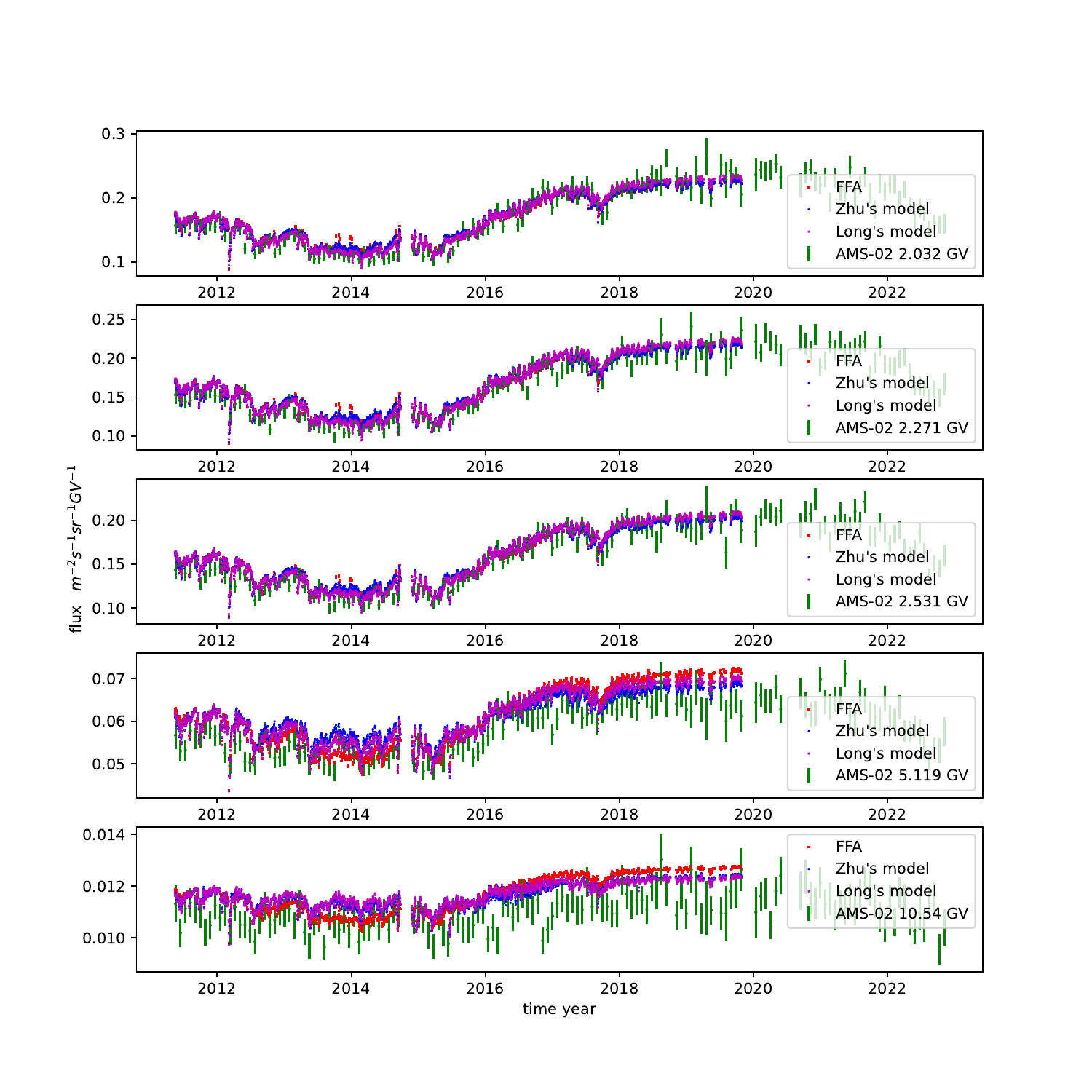}
    \caption{FFA model (red points), Zhu's model (blue points) and Long's model(magenta)  prediction of Be fluxes comparing to the AMS-02 data (green points data) at Rigidities = 2.032 GV, 2.271 GV, 2.531 GV and 5.119 GV and 12.49 GV (from top to bottom).  }
    \label{fig:Be}
\end{figure*}

\begin{figure*}
    \centering
    \includegraphics[scale=0.7]{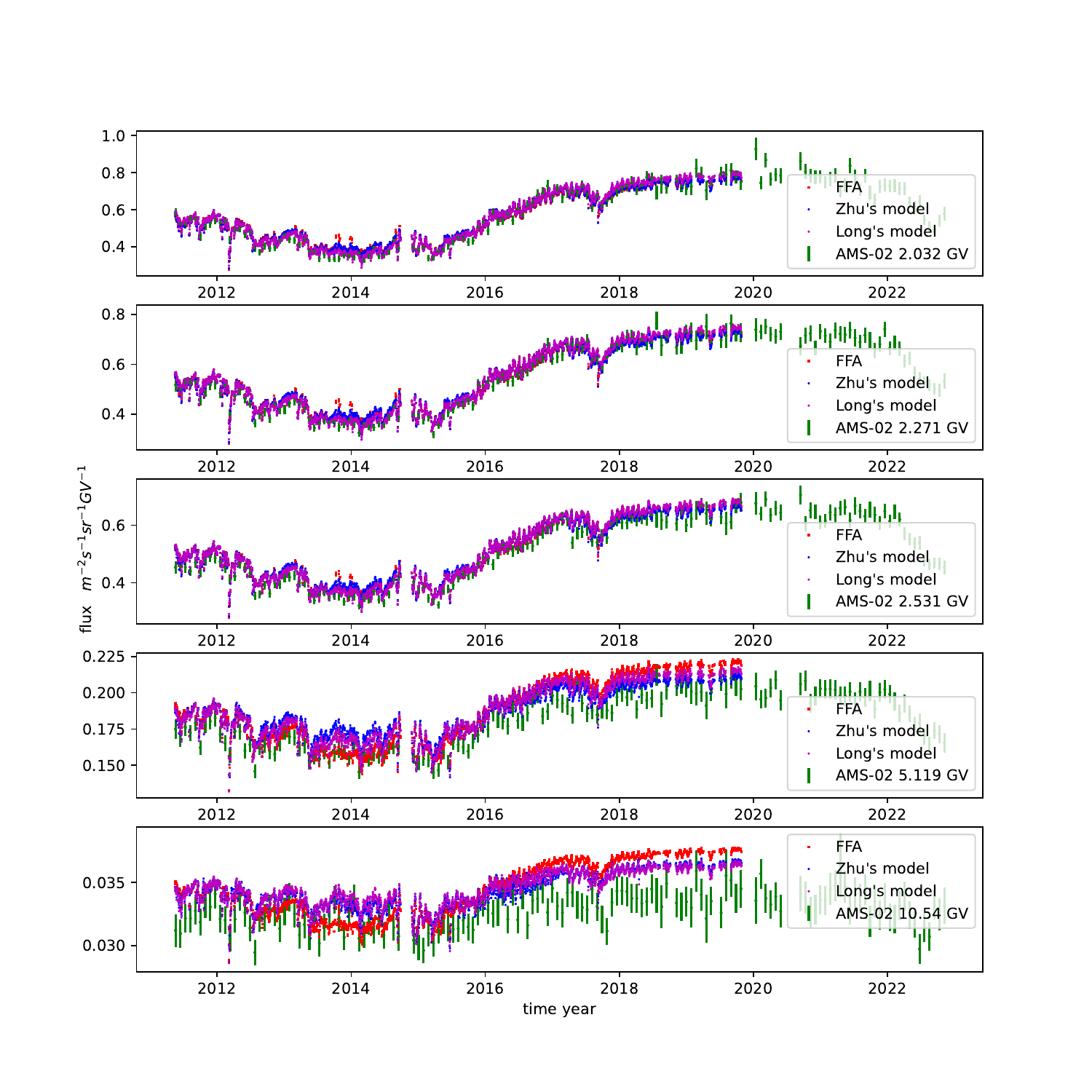}
    \caption{FFA model (red points), Zhu's model (blue points) and Long's model(magenta)  prediction of B fluxes comparing to the AMS-02 data (green points data) at Rigidities = 2.032 GV, 2.271 GV, 2.531 GV and 5.119 GV and 12.49 GV (from top to bottom).  }
    \label{fig:B}
\end{figure*}

\begin{figure*}
    \centering
    \includegraphics[scale=0.7]{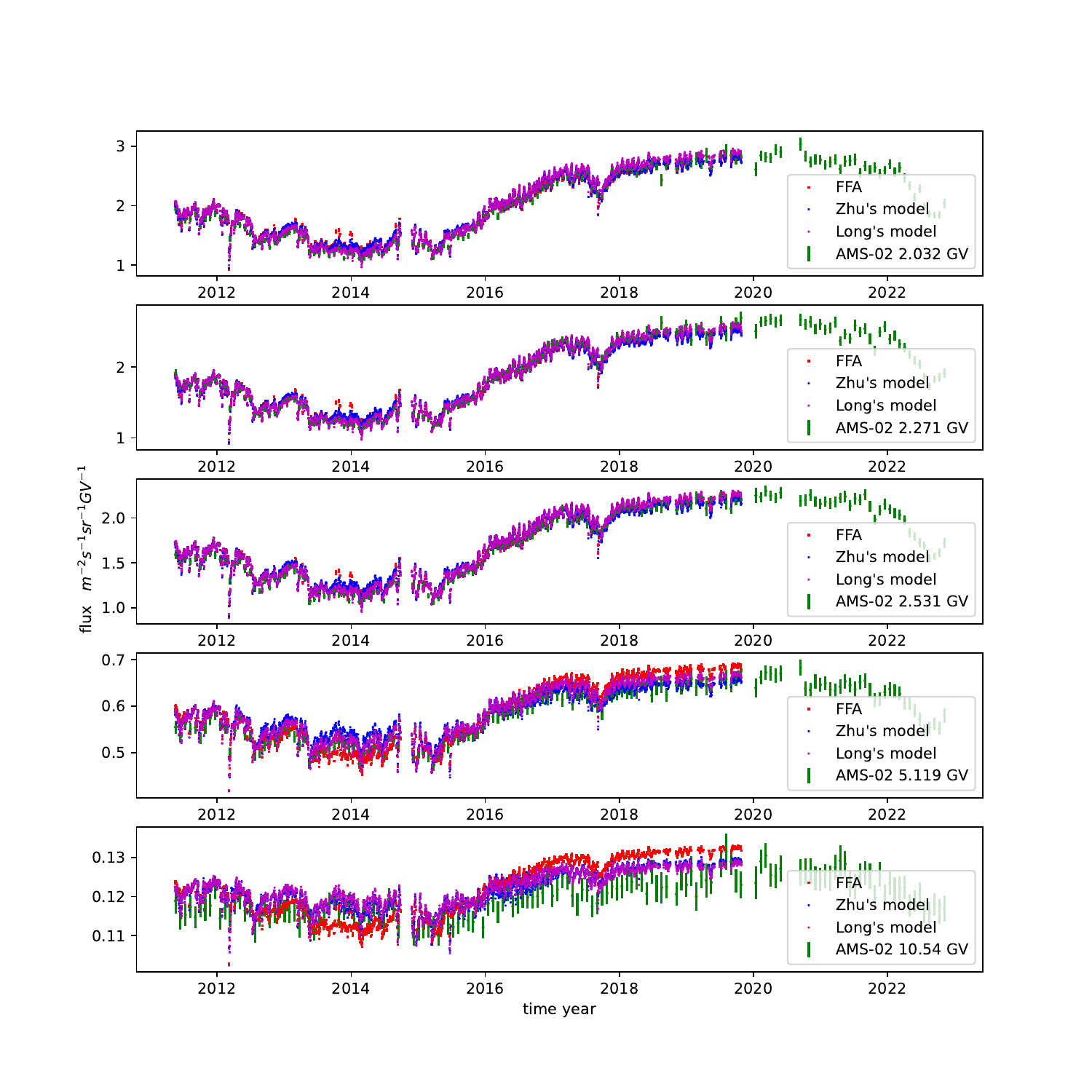}
    \caption{FFA model (red points), Zhu's model (blue points) and Long's model(magenta)  prediction of C fluxes comparing to the AMS-02 data (green points data) at Rigidities = 2.032 GV, 2.271 GV, 2.531 GV and 5.119 GV and 12.49 GV (from top to bottom).  }
    \label{fig:C}
\end{figure*}

\begin{figure*}
    \centering
    \includegraphics[scale=0.7]{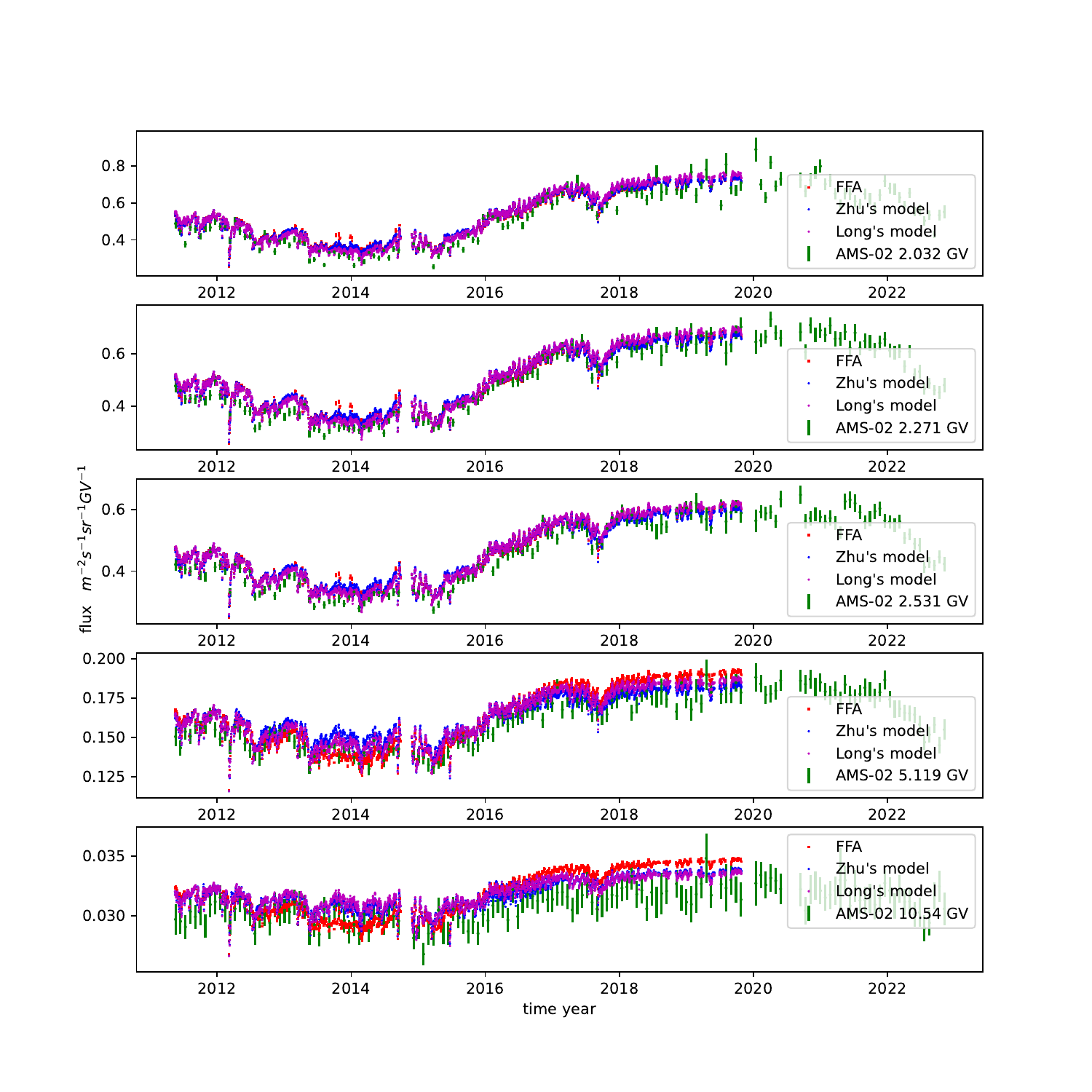}
    \caption{FFA model (red points), Zhu's model (blue points) and Long's model(magenta)  prediction of N fluxes comparing to the AMS-02 data (green points data) at Rigidities = 2.032 GV, 2.271 GV, 2.531 GV and 5.119 GV and 12.49 GV (from top to bottom).  }
    \label{fig:N}
\end{figure*}

\begin{figure*}
    \centering
    \includegraphics[scale=0.7]{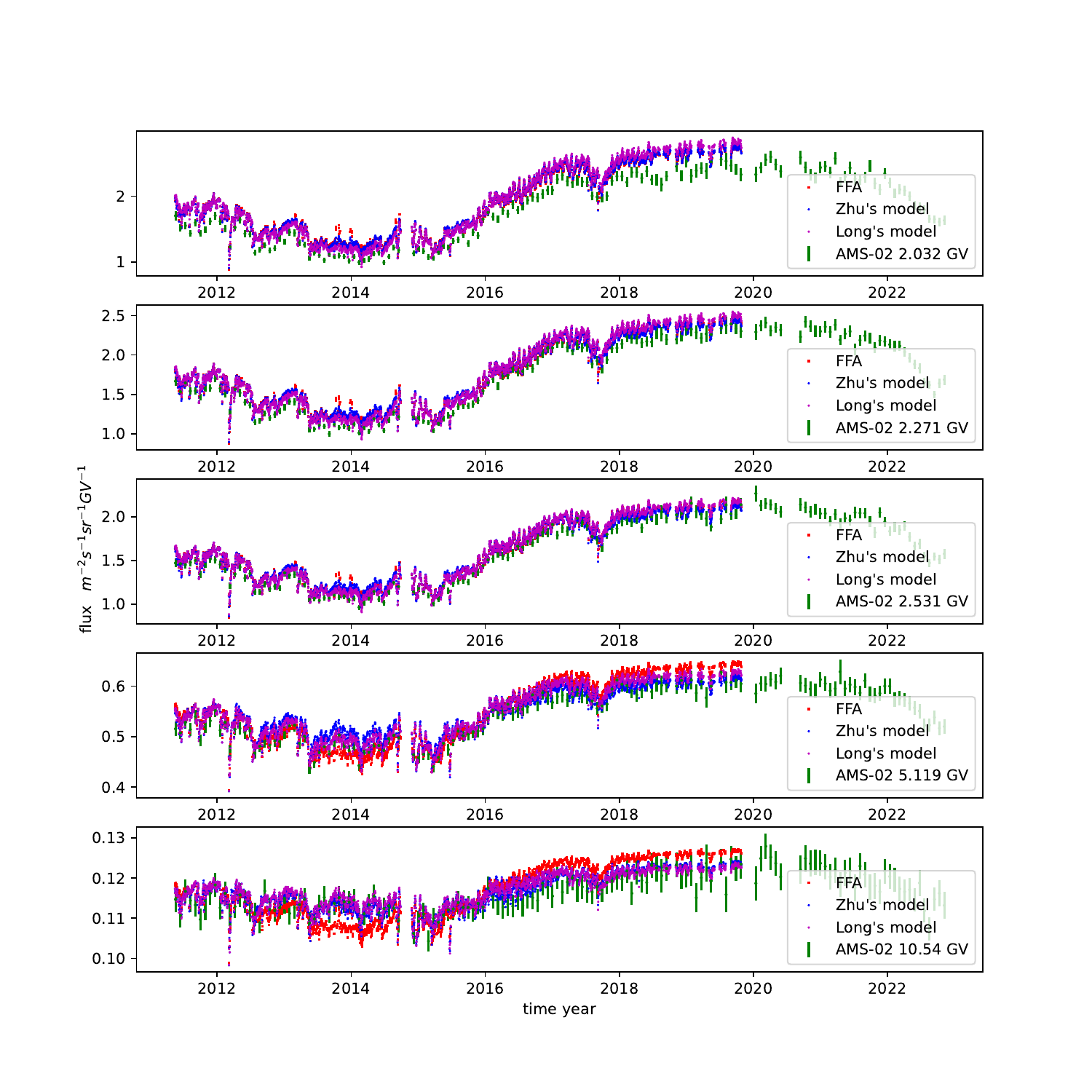}
    \caption{FFA model (red points), Zhu's model (blue points) and Long's model(magenta)  prediction of O fluxes comparing to the AMS-02 data (green points data) at Rigidities = 2.032 GV, 2.271 GV, 2.531 GV and 5.119 GV and 12.49 GV (from top to bottom).   }
    \label{fig:O}
\end{figure*}

\zcr{We present the predicted daily fluxes for Li, Be, B, C, N, and O at selected rigidities (2.032 GV, 2.271 GV, 2.531 GV, 5.119 GV, and 12.49 GV), alongside the corresponding AMS-02 measurements \citep{PhysRevLett.134.051001} }
in Figures \ref{fig:Li}, \ref{fig:Be}, \ref{fig:B}, \ref{fig:C}, \ref{fig:N}, \ref{fig:O}.
For  most cases, the predictions agree with the measurements within their 1 $\sim$ 2 $\sigma$ confidence interval except for O fluxes at low rigidities. The model prediction of O fluxes is a little higher than the measurement by 1.1 times at 2.032 GV. This is mainly caused by the over prediction of LIS for O at low rigidities. The AMS-02 data used to obtain the O LIS  from \cite{AGUILAR20211} are higher than the mean flux from Bartels rotation 2426 to 2520 reported in \cite{PhysRevLett.134.051001} at low rigidities.

\begin{figure*}
    \centering
    \includegraphics[scale=0.7]{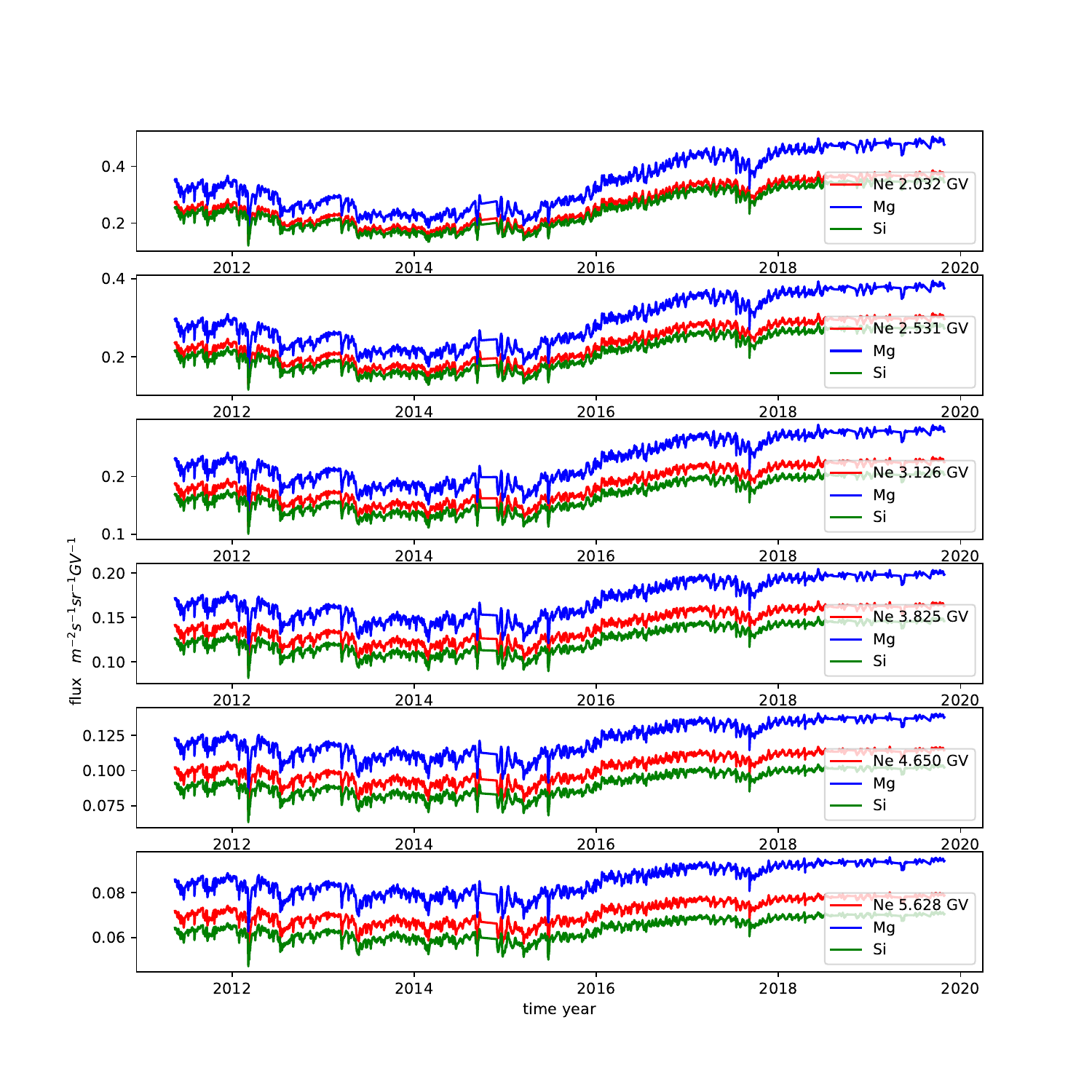}
    \caption{Zhu's model prediction  of Ne (red lines), Mg (blue lines) and Si (green lines) fluxes  at Rigidities = 2.032 GV, 2.531GV, 3.126 GV, 3.825 GV, 4.650 GV and 5.628 GV (from top to bottom).  }
    \label{fig:NeMgSi}
\end{figure*}

\begin{figure*}
    \centering
    \includegraphics[scale=0.7]{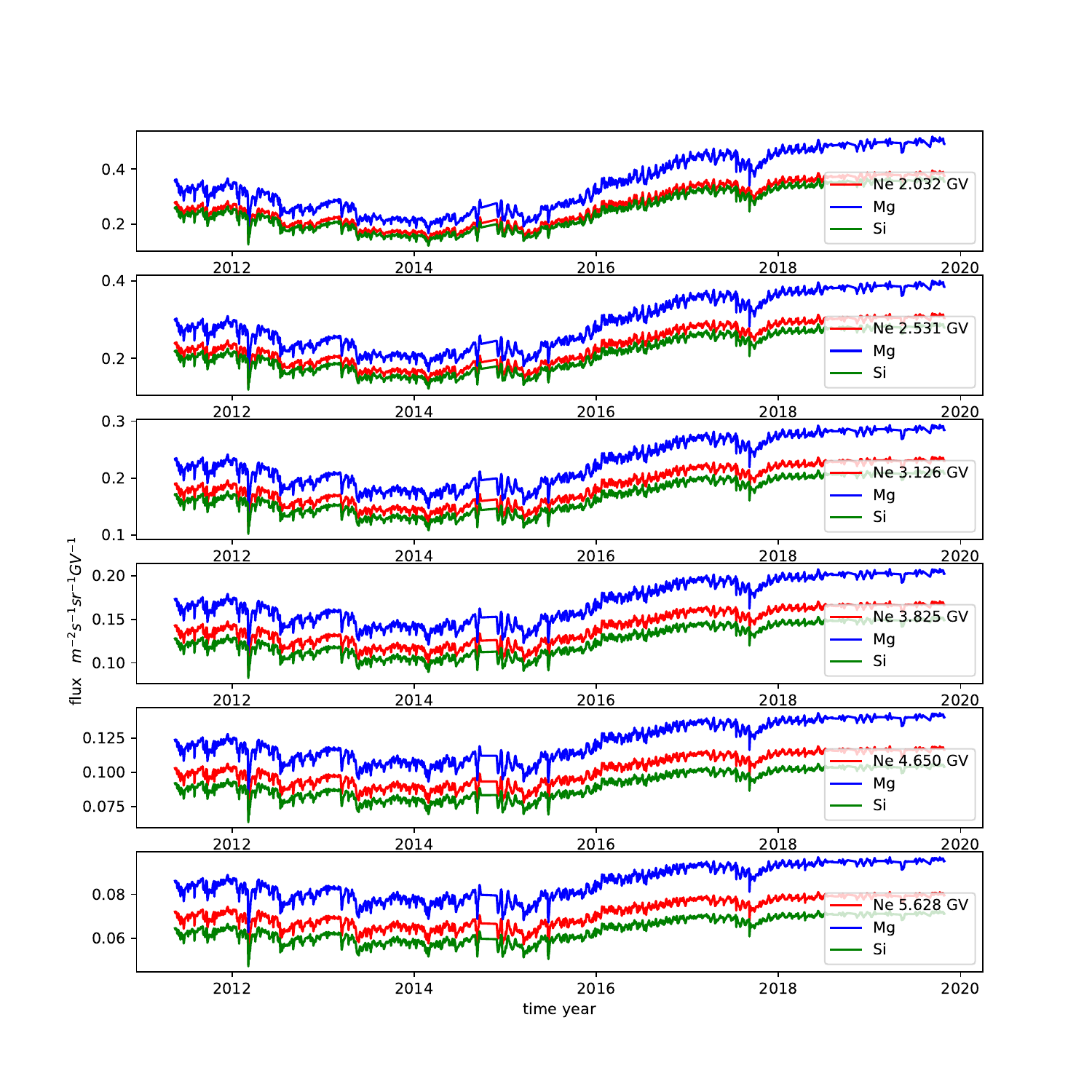}
    \caption{Long's model prediction  of Ne (red lines), Mg (blue lines) and Si (green lines) fluxes  at Rigidities = 2.032 GV, 2.531GV, 3.126 GV, 3.825 GV, 4.650 GV and 5.628 GV (from top to bottom).  }
    \label{fig:NeMgSi2}
\end{figure*}

In Figure \ref{fig:NeMgSi} and Figure \ref{fig:NeMgSi2} we show the Zhu and Long's model prediction daily fluxes of Ne, Mg and Si at selected rigidities. More data at rigidities from 2 to 60 GV can be found in our homepage. Similar to the other GCRs in this work, the fluxes show similar He but not identical time behavior. There are some long-term variations in the flux ratios, and they show opposite trends at low and high rigidities for some GCRs  in the model prediction. However, these variations are too subtle to be detected by the AMS-02.  \zcrr{The daily flux predictions for Li, Be, B, C, N, O, Ne, Mg, and Si generated in this study from 2 to 60 GV  are publicly available at the data repository
 \dataset[doi:10.5281/zenodo.21813652]{https://doi.org/10.5281/zenodo.21813652}.} We can also predict the monthly fluxes of Ne, Mg, and Si on the basis of the monthly fluxes of p, He, and other GCRs. This prediction work will be carried out in the future.

\section{Conclusion}

The precision daily measurements of GCRs spectra by the AMS-02 offer an unprecedented opportunity to probe the dynamics of solar modulation. In this study, we derive the LIS for lithium, beryllium, boron, carbon, nitrogen, oxygen, neon, magnesium, and silicon (Li–Si) by synthesizing Voyager 2 data at low rigidities and AMS-02 observations at higher energies. This reconstruction assumes that all positively charged nuclei share solar modulation parameters identical to those of protons (p) and helium (He), as established in our prior work (\cite{zhu2024}). Crucially, ACE data are excluded from the LIS determination due to the solar modulation potential of them is very different to the AMS-02 as the solar modulation potential is rigidity dependent.

Building on the rigidity-dependent solar modulation potential model from \cite{zhu2024,Zhu_2025}, we predict daily fluxes for Li–Si nuclei under the paradigm that all positively charged GCRs experience identical modulation parameters. This assumption is validated by the excellent agreement between the predicted and observed fluxes of Li, Be, B, C, N, and O, which fall within the 1$\sim$2 $\sigma$ confidence intervals over a range of rigidities for the majority of time intervals. Notably, the predicted neon, magnesium, and silicon daily fluxes will be useful for the detection of GCRs in the future. Based on the same principle, method, and process, we can also infer the daily/monthly fluxes  of cosmic rays such as F, Na, S, Al, and Fe. This will be carried out in the future.



\begin{acknowledgments}
Thanks to Yi-Zhong Fan for very helpful discussions. This work is supported by the National Natural Science Foundation of China  (No. 12203103). K.K.D. is supported by the National Key Research and Development Program of China (No. 2022YFF0503304). C.R.Z. is also supported by the Doctoral research start-up funding of Anhui Normal University. T.H.L. and Y.H.Z. acknowledge the support of the undergraduate Innovation and Entrepreneurship Training Program (No. X202510370575). We acknowledge the use of  data from the \href{https://ams02.space/publications/}{AMS Publications (https://ams02.space/publications/)} and  \href{https://tools.ssdc.asi.it/CosmicRays/}{Cosmic-Ray Database (https://tools.ssdc.asi.it/CosmicRays/)} \citep{DiFelice:2017Hm}.

\end{acknowledgments}

\clearpage

\bibliographystyle{aasjournal}
\bibliography{refs}

@book{MCMC,
  title={Markov Chain Monte Carlo: Stochastic Simulation for Bayesian Inference},
  author={D. Gamerman},
  year={1997},
  publisher={ London: Chapman and Hall }
}

@ARTICLE{2017PhRvD..95h3007Y,
   author = {{Yuan}, Q. and {Lin}, S.-J. and {Fang}, K. and {Bi}, X.-J.},
    title = "{Propagation of cosmic rays in the AMS-02 era}",
  journal = {\prd},
archivePrefix = "arXiv",
   eprint = {1701.06149},
 primaryClass = "astro-ph.HE",
     year = 2017,
    month = apr,
   volume = 95,
   number = 08,
      eid = {083007},
    pages = {083007},
      doi = {10.1103/PhysRevD.95.083007},
   adsurl = {http://adsabs.harvard.edu/abs/2017PhRvD..95h3007Y}
}

@ARTICLE{2013Sci...341..150S,
   author = {{Stone}, E.~C. and {Cummings}, A.~C. and {McDonald}, F.~B. and
        {Heikkila}, B.~C. and {Lal}, N. and {Webber}, W.~R.},
    title = "{Voyager 1 Observes Low-Energy Galactic Cosmic Rays in a Region Depleted of Heliospheric Ions}",
  journal = {Science},
     year = 2013,
    month = jul,
   volume = 341,
    pages = {150-153},
      doi = {10.1126/science.1236408},
   adsurl = {http://adsabs.harvard.edu/abs/2013Sci...341..150S}
}

@ARTICLE{2016A&A...591A..94G,
   author = {{Ghelfi}, A. and {Barao}, F. and {Derome}, L. and {Maurin}, D.
	},
    title = "{Non-parametric determination of H and He interstellar fluxes from cosmic-ray data}",
  journal = {\aap},
archivePrefix = "arXiv",
   eprint = {1511.08650},
 primaryClass = "astro-ph.HE",
     year = 2016,
    month = jun,
   volume = 591,
      eid = {A94},
    pages = {A94},
      doi = {10.1051/0004-6361/201527852},
   adsurl = {http://adsabs.harvard.edu/abs/2016A%26A...591A..94G}
}

@ARTICLE{2017AdSpR..60..833G,
   author = {{Ghelfi}, A. and {Maurin}, D. and {Cheminet}, A. and {Derome}, L. and 
	{Hubert}, G. and {Melot}, F.},
    title = "{Neutron monitors and muon detectors for solar modulation studies: 2. {\phiv} time series}",
  journal = {Advances in Space Research},
archivePrefix = "arXiv",
   eprint = {1607.01976},
 primaryClass = "astro-ph.HE",
     year = 2017,
    month = aug,
   volume = 60,
    pages = {833-847},
      doi = {10.1016/j.asr.2016.06.027},
   adsurl = {http://adsabs.harvard.edu/abs/2017AdSpR..60..833G}
}

@ARTICLE{2002PhRvD..66j3511L,
   author = {{Lewis}, A. and {Bridle}, S.},
    title = "{Cosmological parameters from CMB and other data: A Monte Carlo approach}",
  journal = {\prd},
   eprint = {astro-ph/0205436},
     year = 2002,
    month = nov,
   volume = 66,
   number = 10,
    pages = {103511},
      doi = {10.1103/PhysRevD.66.103511},
   adsurl = {http://adsabs.harvard.edu/abs/2002PhRvD..66j3511L}
}

@ARTICLE{2011JGRA..116.2104U,
   author = {{Usoskin}, I.~G. and {Bazilevskaya}, G.~A. and {Kovaltsov}, G.~A.
	},
    title = "{Solar modulation parameter for cosmic rays since 1936 reconstructed from ground-based neutron monitors and ionization chambers}",
  journal = {Journal of Geophysical Research (Space Physics)},
     year = 2011,
    month = feb,
   volume = 116,
      eid = {A02104},
    pages = {A02104},
      doi = {10.1029/2010JA016105},
   adsurl = {http://adsabs.harvard.edu/abs/2011JGRA..116.2104U}
}

@ARTICLE{1965P&SS...13....9P,
   author = {{Parker}, E.~N.},
    title = "{The passage of energetic charged particles through interplanetary space}",
  journal = {\planss},
     year = 1965,
    month = jan,
   volume = 13,
    pages = {9-49},
      doi = {10.1016/0032-0633(65)90131-5},
   adsurl = {http://adsabs.harvard.edu/abs/1965P%26SS...13....9P}
}

@ARTICLE{1968ApJ...154.1011G,
   author = {{Gleeson}, L.~J. and {Axford}, W.~I.},
    title = "{Solar Modulation of Galactic Cosmic Rays}",
  journal = {\apj},
     year = 1968,
    month = dec,
   volume = 154,
    pages = {1011},
      doi = {10.1086/149822},
   adsurl = {http://adsabs.harvard.edu/abs/1968ApJ...154.1011G}
}

@ARTICLE{1967ApJ...149L.115G,
   author = {{Gleeson}, L.~J. and {Axford}, W.~I.},
    title = "{Cosmic Rays in the Interplanetary Medium}",
  journal = {\apjl},
     year = 1967,
    month = sep,
   volume = 149,
    pages = {L115},
      doi = {10.1086/180070},
   adsurl = {http://adsabs.harvard.edu/abs/1967ApJ...149L.115G}
}

@ARTICLE{2014A&A...566A.142Y,
   author = {{Yang}, R.-Z. and {de O{\~n}a Wilhelmi}, E. and {Aharonian}, F.
	},
    title = "{Probing cosmic rays in nearby giant molecular clouds with the Fermi Large Area Telescope}",
  journal = {\aap},
archivePrefix = "arXiv",
   eprint = {1303.7323},
 primaryClass = "astro-ph.HE",
     year = 2014,
    month = jun,
   volume = 566,
      eid = {A142},
    pages = {A142},
      doi = {10.1051/0004-6361/201321044},
   adsurl = {http://adsabs.harvard.edu/abs/2014A%26A...566A.142Y}
}

@article{Liu_2012,
   title={Cosmic ray Monte Carlo: A global fitting method in studying the properties of the new sources of cosmic<mml:math xmlns:mml="http://www.w3.org/1998/Math/MathML" display="inline"><mml:msup><mml:mi>e</mml:mi><mml:mo>±</mml:mo></mml:msup></mml:math>excesses},
   volume={85},
   ISSN={1550-2368},
   url={http://dx.doi.org/10.1103/PhysRevD.85.043507},
   DOI={10.1103/physrevd.85.043507},
   number={4},
   journal={Physical Review D},
   publisher={American Physical Society (APS)},
   author={Liu, Jie and Yuan, Qiang and Bi, Xiao-Jun and Li, Hong and Zhang, Xinmin},
   year={2012},
   month=feb }

@article{USOSKIN20152940,
title = {Force-field parameterization of the galactic cosmic ray spectrum: Validation for Forbush decreases},
journal = {Advances in Space Research},
volume = {55},
number = {12},
pages = {2940-2945},
year = {2015},
issn = {0273-1177},
doi = {https://doi.org/10.1016/j.asr.2015.03.009},
url = {https://www.sciencedirect.com/science/article/pii/S0273117715001921},
author = {I.G. Usoskin and G.A. Kovaltsov and O. Adriani and G.C. Barbarino and G.A. Bazilevskaya and R. Bellotti and M. Boezio and E.A. Bogomolov and M. Bongi and V. Bonvicini and S. Bottai and A. Bruno and F. Cafagna and D. Campana and R. Carbone and P. Carlson and M. Casolino and G. Castellini and C. {De Donato} and C. {De Santis} and N. {De Simone} and V. {Di Felice} and V. Formato and A.M. Galper and A.V. Karelin and S.V. Koldashov and S. Koldobskiy and S.Y. Krutkov and A.N. Kvashnin and A. Leonov and V. Malakhov and L. Marcelli and M. Martucci and A.G. Mayorov and W. Menn and M. Mergé and V.V. Mikhailov and E. Mocchiutti and A. Monaco and N. Mori and R. Munini and G. Osteria and F. Palma and B. Panico and P. Papini and M. Pearce and P. Picozza and C. Pizzolotto and M. Ricci and S.B. Ricciarini and L. Rossetto and R. Sarkar and V. Scotti and M. Simon and R. Sparvoli and P. Spillantini and Y.I. Stozhkov and A. Vacchi and E. Vannuccini and G.I. Vasilyev and S.A. Voronov and Y.T. Yurkin and G. Zampa and N. Zampa and V.G. Zverev}
}

@ARTICLE{2016ApJ...829....8C,
   author = {{Corti}, C. and {Bindi}, V. and {Consolandi}, C. and {Whitman}, K.
	},
    title = "{Solar Modulation of the Local Interstellar Spectrum with Voyager 1, AMS-02, PAMELA, and BESS}",
  journal = {\apj},
archivePrefix = "arXiv",
   eprint = {1511.08790},
 primaryClass = "astro-ph.HE",
     year = 2016,
    month = sep,
   volume = 829,
      eid = {8},
    pages = {8},
      doi = {10.3847/0004-637X/829/1/8},
   adsurl = {http://adsabs.harvard.edu/abs/2016ApJ...829....8C}
}

@ARTICLE{1998ApJ...493..694M,
   author = {{Moskalenko}, I.~V. and {Strong}, A.~W.},
    title = "{Production and Propagation of Cosmic-Ray Positrons and Electrons}",
  journal = {\apj},
   eprint = {astro-ph/9710124},
     year = 1998,
    month = jan,
   volume = 493,
    pages = {694-707},
      doi = {10.1086/305152},
   adsurl = {https://ui.adsabs.harvard.edu/abs/1998ApJ...493..694M}
}

@article{Zhu:2018jbk,
      author         = "Zhu, Cheng-Rui and Yuan, Qiang and Wei, Da-Ming",
      title          = "{Studies on cosmic ray nuclei with Voyager, ACE and
                        AMS-02: I. local interstellar spectra and solar
                        modulation}",
      journal        = "Astrophys. J.",
      volume         = "863",
      year           = "2018",
      number         = "2",
      pages          = "119",
      doi            = "10.3847/1538-4357/aacff9",
      eprint         = "1807.09470",
      archivePrefix  = "arXiv",
      primaryClass   = "astro-ph.HE",
      SLACcitation   = "%%CITATION = ARXIV:1807.09470;%%"
}

@article{PhysRevLett.121.051101,
  title = {Observation of Fine Time Structures in the Cosmic Proton and Helium Fluxes with the Alpha Magnetic Spectrometer on the International Space Station},
  author = {Aguilar, M. and Ali Cavasonza, L. and Alpat, B. and Ambrosi, G. and Arruda, L. and Attig, N. and Aupetit, S. and Azzarello, P. and Bachlechner, A. and Barao, F. and Barrau, A. and Barrin, L. and Bartoloni, A. and Basara, L. and Ba\ifmmode \mbox{\c{s}}\else \c{s}\fi{}e\ifmmode \breve{g}\else \u{g}\fi{}mez-du Pree, S. and Battarbee, M. and Battiston, R. and Becker, U. and Behlmann, M. and Beischer, B. and Berdugo, J. and Bertucci, B. and Bindel, K. F. and Bindi, V. and de Boer, W. and Bollweg, K. and Bonnivard, V. and Borgia, B. and Boschini, M. J. and Bourquin, M. and Bueno, E. F. and Burger, J. and Cadoux, F. and Cai, X. D. and Capell, M. and Caroff, S. and Casaus, J. and Castellini, G. and Cervelli, F. and Chae, M. J. and Chang, Y. H. and Chen, A. I. and Chen, G. M. and Chen, H. S. and Chen, Y. and Cheng, L. and Chou, H. Y. and Choumilov, E. and Choutko, V. and Chung, C. H. and Clark, C. and Clavero, R. and Coignet, G. and Consolandi, C. and Contin, A. and Corti, C. and Creus, W. and Crispoltoni, M. and Cui, Z. and Dadzie, K. and Dai, Y. M. and Datta, A. and Delgado, C. and Della Torre, S. and Demirk\"oz, M. B. and Derome, L. and Di Falco, S. and Dimiccoli, F. and D\'{\i}az, C. and von Doetinchem, P. and Dong, F. and Donnini, F. and Duranti, M. and D'Urso, D. and Egorov, A. and Eline, A. and Eronen, T. and Feng, J. and Fiandrini, E. and Fisher, P. and Formato, V. and Galaktionov, Y. and Gallucci, G. and Garc\'{\i}a-L\'opez, R. J. and Gargiulo, C. and Gast, H. and Gebauer, I. and Gervasi, M. and Ghelfi, A. and Giovacchini, F. and G\'omez-Coral, D. M. and Gong, J. and Goy, C. and Grabski, V. and Grandi, D. and Graziani, M. and Guo, K. H. and Haino, S. and Han, K. C. and He, Z. H. and Heil, M. and Hoffman, J. and Hsieh, T. H. and Huang, H. and Huang, Z. C. and Huh, C. and Incagli, M. and Ionica, M. and Jang, W. Y. and Jia, Yi and Jinchi, H. and Kang, S. C. and Kanishev, K. and Khiali, B. and Kim, G. N. and Kim, K. S. and Kirn, Th. and Konak, C. and Kounina, O. and Kounine, A. and Koutsenko, V. and Kulemzin, A. and La Vacca, G. and Laudi, E. and Laurenti, G. and Lazzizzera, I. and Lebedev, A. and Lee, H. T. and Lee, S. C. and Leluc, C. and Li, H. S. and Li, J. Q. and Li, Q. and Li, T. X. and Li, Z. H. and Li, Z. Y. and Light, C. and Lim, S. and Lin, C. H. and Lipari, P. and Lippert, T. and Liu, D. and Liu, Hu and Lordello, V. D. and Lu, S. Q. and Lu, Y. S. and Luebelsmeyer, K. and Luo, F. and Luo, J. Z. and Luo, X. and Lyu, S. S. and Machate, F. and Ma\~n\'a, C. and Mar\'{\i}n, J. and Martin, T. and Mart\'{\i}nez, G. and Masi, N. and Maurin, D. and Menchaca-Rocha, A. and Meng, Q. and Mikuni, V. M. and Mo, D. C. and Mott, P. and Nelson, T. and Ni, J. Q. and Nikonov, N. and Nozzoli, F. and Oliva, A. and Orcinha, M. and Palermo, M. and Palmonari, F. and Palomares, C. and Paniccia, M. and Pauluzzi, M. and Pensotti, S. and Perrina, C. and Phan, H. D. and Picot-Clemente, N. and Pilo, F. and Pizzolotto, C. and Plyaskin, V. and Pohl, M. and Poireau, V. and Popkow, A. and Quadrani, L. and Qi, X. M. and Qin, X. and Qu, Z. Y. and R\"aih\"a, T. and Rancoita, P. G. and Rapin, D. and Ricol, J. S. and Rosier-Lees, S. and Rozhkov, A. and Rozza, D. and Sagdeev, R. and Schael, S. and Schmidt, S. M. and Schulz von Dratzig, A. and Schwering, G. and Seo, E. S. and Shan, B. S. and Shi, J. Y. and Siedenburg, T. and Son, D. and Song, J. W. and Tacconi, M. and Tang, X. W. and Tang, Z. C. and Tescaro, D. and Ting, Samuel C. C. and Ting, S. M. and Tomassetti, N. and Torsti, J. and T\"urko\ifmmode \breve{g}\else \u{g}\fi{}lu, C. and Urban, T. and Vagelli, V. and Valente, E. and Valtonen, E. and V\'azquez Acosta, M. and Vecchi, M. and Velasco, M. and Vialle, J. P. and Wang, L. Q. and Wang, N. H. and Wang, Q. L. and Wang, X. and Wang, X. Q. and Wang, Z. X. and Wei, C. C. and Weng, Z. L. and Whitman, K. and Wu, H. and Wu, X. and Xiong, R. Q. and Xu, W. and Yan, Q. and Yang, J. and Yang, M. and Yang, Y. and Yi, H. and Yu, Y. J. and Yu, Z. Q. and Zannoni, M. and Zeissler, S. and Zhang, C. and Zhang, F. and Zhang, J. and Zhang, J. H. and Zhang, S. W. and Zhang, Z. and Zheng, Z. M. and Zhuang, H. L. and Zhukov, V. and Zichichi, A. and Zimmermann, N. and Zuccon, P.},
  collaboration = {AMS Collaboration},
  journal = {Phys. Rev. Lett.},
  volume = {121},
  issue = {5},
  pages = {051101},
  numpages = {7},
  year = {2018},
  month = {Jul},
  publisher = {American Physical Society},
  doi = {10.1103/PhysRevLett.121.051101},
  url = {https://link.aps.org/doi/10.1103/PhysRevLett.121.051101}
}

@article{PhysRevLett.121.051102,
  title = {Observation of Complex Time Structures in the Cosmic-Ray Electron and Positron Fluxes with the Alpha Magnetic Spectrometer on the International Space Station},
  author = {Aguilar, M. and Cavasonza, L. Ali and Ambrosi, G. and Arruda, L. and Attig, N. and Aupetit, S. and Azzarello, P. and Bachlechner, A. and Barao, F. and Barrau, A. and Barrin, L. and Bartoloni, A. and Basara, L. and Ba\ifmmode \mbox{\c{s}}\else \c{s}\fi{}e\ifmmode \breve{g}\else \u{g}\fi{}mez-du Pree, S. and Battarbee, M. and Battiston, R. and Becker, U. and Behlmann, M. and Beischer, B. and Berdugo, J. and Bertucci, B. and Bindel, K. F. and Bindi, V. and de Boer, W. and Bollweg, K. and Bonnivard, V. and Borgia, B. and Boschini, M. J. and Bourquin, M. and Bueno, E. F. and Burger, J. and Cadoux, F. and Cai, X. D. and Capell, M. and Caroff, S. and Casaus, J. and Castellini, G. and Cervelli, F. and Chae, M. J. and Chang, Y. H. and Chen, A. I. and Chen, G. M. and Chen, H. S. and Chen, Y. and Cheng, L. and Chou, H. Y. and Choumilov, E. and Choutko, V. and Chung, C. H. and Clark, C. and Clavero, R. and Coignet, G. and Consolandi, C. and Contin, A. and Corti, C. and Creus, W. and Crispoltoni, M. and Cui, Z. and Dadzie, K. and Dai, Y. M. and Datta, A. and Delgado, C. and Della Torre, S. and Demirk\"oz, M. B. and Derome, L. and Di Falco, S. and Dimiccoli, F. and D\'{\i}az, C. and von Doetinchem, P. and Dong, F. and Donnini, F. and Duranti, M. and D'Urso, D. and Egorov, A. and Eline, A. and Eronen, T. and Feng, J. and Fiandrini, E. and Fisher, P. and Formato, V. and Galaktionov, Y. and Gallucci, G. and Garc\'{\i}a-L\'opez, R. J. and Gargiulo, C. and Gast, H. and Gebauer, I. and Gervasi, M. and Ghelfi, A. and Giovacchini, F. and G\'omez-Coral, D. M. and Gong, J. and Goy, C. and Grabski, V. and Grandi, D. and Graziani, M. and Guo, K. H. and Haino, S. and Han, K. C. and He, Z. H. and Heil, M. and Hsieh, T. H. and Huang, H. and Huang, Z. C. and Huh, C. and Incagli, M. and Ionica, M. and Jang, W. Y. and Jia, Yi and Jinchi, H. and Kang, S. C. and Kanishev, K. and Khiali, B. and Kim, G. N. and Kim, K. S. and Kirn, Th. and Konak, C. and Kounina, O. and Kounine, A. and Koutsenko, V. and Kulemzin, A. and La Vacca, G. and Laudi, E. and Laurenti, G. and Lazzizzera, I. and Lebedev, A. and Lee, H. T. and Lee, S. C. and Leluc, C. and Li, H. S. and Li, J. Q. and Li, Q. and Li, T. X. and Li, Z. H. and Li, Z. Y. and Lim, S. and Lin, C. H. and Lipari, P. and Lippert, T. and Liu, D. and Liu, Hu and Lordello, V. D. and Lu, S. Q. and Lu, Y. S. and Luebelsmeyer, K. and Luo, F. and Luo, J. Z. and Lyu, S. S. and Machate, F. and Ma\~n\'a, C. and Mar\'{\i}n, J. and Martin, T. and Mart\'{\i}nez, G. and Masi, N. and Maurin, D. and Menchaca-Rocha, A. and Meng, Q. and Mikuni, V. M. and Mo, D. C. and Mott, P. and Nelson, T. and Ni, J. Q. and Nikonov, N. and Nozzoli, F. and Oliva, A. and Orcinha, M. and Palermo, M. and Palmonari, F. and Palomares, C. and Paniccia, M. and Pauluzzi, M. and Pensotti, S. and Perrina, C. and Phan, H. D. and Picot-Clemente, N. and Pilo, F. and Pizzolotto, C. and Plyaskin, V. and Pohl, M. and Poireau, V. and Quadrani, L. and Qi, X. M. and Qin, X. and Qu, Z. Y. and R\"aih\"a, T. and Rancoita, P. G. and Rapin, D. and Ricol, J. S. and Rosier-Lees, S. and Rozhkov, A. and Rozza, D. and Sagdeev, R. and Schael, S. and Schmidt, S. M. and von Dratzig, A. Schulz and Schwering, G. and Seo, E. S. and Shan, B. S. and Shi, J. Y. and Siedenburg, T. and Son, D. and Song, J. W. and Tacconi, M. and Tang, X. W. and Tang, Z. C. and Tescaro, D. and Ting, Samuel C. C. and Ting, S. M. and Tomassetti, N. and Torsti, J. and T\"urko\ifmmode \breve{g}\else \u{g}\fi{}lu, C. and Urban, T. and Vagelli, V. and Valente, E. and Valtonen, E. and V\'azquez Acosta, M. and Vecchi, M. and Velasco, M. and Vialle, J. P. and Wang, L. Q. and Wang, N. H. and Wang, Q. L. and Wang, X. and Wang, X. Q. and Wang, Z. X. and Wei, C. C. and Weng, Z. L. and Whitman, K. and Wu, H. and Wu, X. and Xiong, R. Q. and Xu, W. and Yan, Q. and Yang, J. and Yang, M. and Yang, Y. and Yi, H. and Yu, Y. J. and Yu, Z. Q. and Zannoni, M. and Zeissler, S. and Zhang, C. and Zhang, F. and Zhang, J. and Zhang, J. H. and Zhang, S. W. and Zhang, Z. and Zheng, Z. M. and Zhuang, H. L. and Zhukov, V. and Zichichi, A. and Zimmermann, N. and Zuccon, P.},
  collaboration = {AMS Collaboration},
  journal = {Phys. Rev. Lett.},
  volume = {121},
  issue = {5},
  pages = {051102},
  numpages = {8},
  year = {2018},
  month = {Jul},
  publisher = {American Physical Society},
  doi = {10.1103/PhysRevLett.121.051102},
  url = {https://link.aps.org/doi/10.1103/PhysRevLett.121.051102}
}

@article{2003JA010098,
author = {Caballero-Lopez, R. A. and Moraal, H.},
title = {Limitations of the force field equation to describe cosmic ray modulation},
journal = {Journal of Geophysical Research: Space Physics},
volume = {109},
number = {A1},
pages = {},
doi = {https://doi.org/10.1029/2003JA010098},
url = {https://agupubs.onlinelibrary.wiley.com/doi/abs/10.1029/2003JA010098},
year = {2004}
}

@article{PhysRevD.106.063021,
  title = {Constraining the charge-, time-, and rigidity-dependence of cosmic-ray solar modulation with AMS-02 observations during Solar Cycle 24},
  author = {Cholis, Ilias and McKinnon, Ian},
  journal = {Phys. Rev. D},
  volume = {106},
  issue = {6},
  pages = {063021},
  numpages = {13},
  year = {2022},
  month = {Sep},
  publisher = {American Physical Society},
  doi = {10.1103/PhysRevD.106.063021},
  url = {https://link.aps.org/doi/10.1103/PhysRevD.106.063021}
}

@Article{Potgieter2013,
author="Potgieter, Marius S.",
title="Solar Modulation of Cosmic Rays",
journal="Living Reviews in Solar Physics",
year="2013",
month="Jun",
day="13",
volume="10",
number="1",
pages="3",
issn="1614-4961",
doi="10.12942/lrsp-2013-3",
url="https://doi.org/10.12942/lrsp-2013-3"
}

@ARTICLE{2011Sci...332...69A,
       author = {{Adriani}, O. and {Barbarino}, G.~C. and {Bazilevskaya}, G.~A. and
         {Bellotti}, R. and {Boezio}, M. and {Bogomolov}, E.~A. and
         {Bonechi}, L. and {Bongi}, M. and {Bonvicini}, V. and {Borisov}, S. and
         {Bottai}, S. and {Bruno}, A. and {Cafagna}, F. and {Campana}, D. and
         {Carbone}, R. and {Carlson}, P. and {Casolino}, M. and
         {Castellini}, G. and {Consiglio}, L. and {De Pascale}, M.~P. and
         {De Santis}, C. and {De Simone}, N. and {Di Felice}, V. and
         {Galper}, A.~M. and {Gillard}, W. and {Grishantseva}, L. and
         {Jerse}, G. and {Karelin}, A.~V. and {Koldashov}, S.~V. and
         {Krutkov}, S.~Y. and {Kvashnin}, A.~N. and {Leonov}, A. and
         {Malakhov}, V. and {Malvezzi}, V. and {Marcelli}, L. and
         {Mayorov}, A.~G. and {Menn}, W. and {Mikhailov}, V.~V. and
         {Mocchiutti}, E. and {Monaco}, A. and {Mori}, N. and {Nikonov}, N. and
         {Osteria}, G. and {Palma}, F. and {Papini}, P. and {Pearce}, M. and
         {Picozza}, P. and {Pizzolotto}, C. and {Ricci}, M. and
         {Ricciarini}, S.~B. and {Rossetto}, L. and {Sarkar}, R. and
         {Simon}, M. and {Sparvoli}, R. and {Spillantini}, P. and
         {Stozhkov}, Y.~I. and {Vacchi}, A. and {Vannuccini}, E. and
         {Vasilyev}, G. and {Voronov}, S.~A. and {Yurkin}, Y.~T. and {Wu}, J. and
         {Zampa}, G. and {Zampa}, N. and {Zverev}, V.~G.},
        title = "{PAMELA Measurements of Cosmic-Ray Proton and Helium Spectra}",
      journal = {Science},
         year = "2011",
        month = "Apr",
       volume = {332},
       number = {6025},
        pages = {69},
          doi = {10.1126/science.1199172},
archivePrefix = {arXiv},
       eprint = {1103.4055},
 primaryClass = {astro-ph.HE},
       adsurl = {https://ui.adsabs.harvard.edu/abs/2011Sci...332...69A}
}

@article{cite-key,
	Author = {Ambrosi, G. and An, Q. and Asfandiyarov, R. and Azzarello, P. and Bernardini, P. and Bertucci, B. and Cai, M. S. and Chang, J. and Chen, D. Y. and Chen, H. F. and Chen, J. L. and Chen, W. and Cui, M. Y. and Cui, T. S. and D'Amone, A. and De Benedittis, A. and De Mitri, I. and Di Santo, M. and Dong, J. N. and Dong, T. K. and Dong, Y. F. and Dong, Z. X. and Donvito, G. and Droz, D. and Duan, K. K. and Duan, J. L. and Duranti, M. and D'Urso, D. and Fan, R. R. and Fan, Y. Z. and Fang, F. and Feng, C. Q. and Feng, L. and Fusco, P. and Gallo, V. and Gan, F. J. and Gao, M. and Gao, S. S. and Gargano, F. and Garrappa, S. and Gong, K. and Gong, Y. Z. and Guo, D. Y. and Guo, J. H. and Hu, Y. M. and Huang, G. S. and Huang, Y. Y. and Ionica, M. and Jiang, D. and Jiang, W. and Jin, X. and Kong, J. and Lei, S. J. and Li, S. and Li, X. and Li, W. L. and Li, Y. and Liang, Y. F. and Liang, Y. M. and Liao, N. H. and Liu, H. and Liu, J. and Liu, S. B. and Liu, W. Q. and Liu, Y. and Loparco, F. and Ma, M. and Ma, P. X. and Ma, S. Y. and Ma, T. and Ma, X. Q. and Ma, X. Y. and Marsella, G. and Mazziotta, M. N. and Mo, D. and Niu, X. Y. and Peng, X. Y. and Peng, W. X. and Qiao, R. and Rao, J. N. and Salinas, M. M. and Shang, G. Z. and H. Shen, W. and Shen, Z. Q. and Shen, Z. T. and Song, J. X. and Su, H. and Su, M. and Sun, Z. Y. and Surdo, A. and Teng, X. J. and Tian, X. B. and Tykhonov, A. and Vagelli, V. and Vitillo, S. and Wang, C. and Wang, H. and Wang, H. Y. and Wang, J. Z. and Wang, L. G. and Wang, Q. and Wang, S. and Wang, X. H. and Wang, X. L. and Wang, Y. F. and Wang, Y. P. and Wang, Y. Z. and Wen, S. C. and Wang, Z. M. and Wei, D. M. and Wei, J. J. and Wei, Y. F. and Wu, D. and Wu, J. and Wu, L. B. and Wu, S. S. and Wu, X. and Xi, K. and Xia, Z. Q. and Xin, Y. L. and Xu, H. T. and Xu, Z. L. and Xu, Z. Z. and Xue, G. F. and Yang, H. B. and Yang, P. and Yang, Y. Q. and Yang, Z. L. and Yao, H. J. and Yu, Y. H. and Yuan, Q. and Yue, C. and Zang, J. J. and Zhang, C. and Zhang, D. L. and Zhang, F. and Zhang, J. B. and Zhang, J. Y. and Zhang, J. Z. and Zhang, L. and Zhang, P. F. and Zhang, S. X. and Zhang, W. Z. and Zhang, Y. and Zhang, Y. J. and Zhang, Y. Q. and Zhang, Y. L. and Zhang, Y. P. and Zhang, Z. and Zhang, Z. Y. and Zhao, H. and Zhao, H. Y. and Zhao, X. F. and Zhou, C. Y. and Zhou, Y. and Zhu, X. and Zhu, Y. and Zimmer, S. and DAMPE Collaboration},
	Da = {2017/12/01},
	Doi = {10.1038/nature24475},
	Id = {Ambrosi2017},
	Isbn = {1476-4687},
	Journal = {Nature},
	Number = {7683},
	Pages = {63--66},
	Title = {Direct detection of a break in the teraelectronvolt cosmic-ray spectrum of electrons and positrons},
	Ty = {JOUR},
	Url = {https://doi.org/10.1038/nature24475},
	Volume = {552},
	Year = {2017}}

@ARTICLE{2017PhRvL.119y1101A,
   author = {{Aguilar}, M. and {Ali Cavasonza}, L. and {Alpat}, B. and {Ambrosi}, G. and
        {Arruda}, L. and {Attig}, N. and {Aupetit}, S. and {Azzarello}, P. and
        {Bachlechner}, A. and {Barao}, F. and et al.},
    title = "{Observation of the Identical Rigidity Dependence of He, C, and O Cosmic Rays at High Rigidities by the Alpha Magnetic Spectrometer on the International Space Station}",
  journal = {\prl},
     year = 2017,
    month = dec,
   volume = 119,
   number = 25,
      eid = {251101},
    pages = {251101},
      doi = {10.1103/PhysRevLett.119.251101},
   adsurl = {http://adsabs.harvard.edu/abs/2017PhRvL.119y1101A}
}

@article{voyager-2,
	Author = {Stone, Edward C. and Cummings, Alan C. and Heikkila, Bryant C. and Lal, Nand},
	Da = {2019/11/01},
	Doi = {10.1038/s41550-019-0928-3},
	Id = {Stone2019},
	Isbn = {2397-3366},
	Journal = {Nature Astronomy},
	Number = {11},
	Pages = {1013--1018},
	Title = {Cosmic ray measurements from Voyager 2 as it crossed into interstellar space},
	Ty = {JOUR},
	Url = {https://doi.org/10.1038/s41550-019-0928-3},
	Volume = {3},
	Year = {2019}}

@article{PhysRevD.109.083009,
  title = {Probing solar modulation analytic models with cosmic ray periodic spectra},
  author = {Long, Wei-Cheng and Wu, Juan},
  journal = {Phys. Rev. D},
  volume = {109},
  issue = {8},
  pages = {083009},
  numpages = {12},
  year = {2024},
  month = {Apr},
  publisher = {American Physical Society},
  doi = {10.1103/PhysRevD.109.083009},
  url = {https://link.aps.org/doi/10.1103/PhysRevD.109.083009}
}

@article{Zhu:2020koq,
    author={{Zhu}, Cheng-Rui and {Yuan}, Qiang and {Wei}, Da-Ming},
    title = "{Local interstellar spectra and solar modulation of cosmic ray electrons and positrons}",
    eprint = "2007.13989",
    archivePrefix = "arXiv",
    primaryClass = "astro-ph.HE",
    doi = "10.1016/j.astropartphys.2020.102495",
    journal = "Astropart. Phys.",
    volume = "124",
    pages = "102495",
    year = "2021"
}

@ARTICLE{2021ApJ...921..109S,
       author = {{Shen}, Zhenning and {Yang}, Hao and {Zuo}, Pingbing and {Qin}, Gang and {Wei}, Fengsi and {Xu}, Xiaojun and {Xie}, Yanqiong},
        title = "{Solar Modulation of Galactic Cosmic-Ray Protons Based on a Modified Force-field Approach}",
      journal = {\apj},
         year = 2021,
        month = nov,
       volume = {921},
       number = {2},
          eid = {109},
        pages = {109},
          doi = {10.3847/1538-4357/ac1fe8},
       adsurl = {https://ui.adsabs.harvard.edu/abs/2021ApJ...921..109S}
}

@ARTICLE{2022PhRvL.129w1101Z,
       author = {{Zhu}, Cheng-Rui and {Cui}, Ming-Yang and {Xia}, Zi-Qing and {Yu}, Zhao-Huan and {Huang}, Xiaoyuan and {Yuan}, Qiang and {Fan}, Yi-Zhong},
        title = "{Explaining the GeV Antiproton Excess, GeV {\ensuremath{\gamma}} -Ray Excess, and W -Boson Mass Anomaly in an Inert Two Higgs Doublet Model}",
      journal = {\prl},
         year = 2022,
        month = dec,
       volume = {129},
       number = {23},
          eid = {231101},
        pages = {231101},
          doi = {10.1103/PhysRevLett.129.231101},
archivePrefix = {arXiv},
       eprint = {2204.03767},
 primaryClass = {astro-ph.HE},
       adsurl = {https://ui.adsabs.harvard.edu/abs/2022PhRvL.129w1101Z}
}

\end{document}